\documentclass[%
aip,
rsi,
superscriptaddress,
reprint,%
]{revtex4-1}

\usepackage{wasysym}  
\usepackage{graphicx}

\newcommand{\ket}[1]{|#1\rangle}

\begin{document}


\title{Design and characterisation of a compact magnetic shield for ultracold atomic gas experiments} 



\author{A. Farolfi}
\author{D. Trypogeorgos}
\affiliation{INO-CNR BEC Center and Dipartimento di Fisica, Universit\`a di Trento, 38123 Povo, Italy}
\affiliation{Trento Institute for Fundamental Physics and Applications, INFN,  38123 Povo, Italy}
\author{G. Colzi}
\author{E. Fava}
\affiliation{INO-CNR BEC Center and Dipartimento di Fisica, Universit\`a di Trento, 38123 Povo, Italy}
\author{G. Lamporesi}
\author{G. Ferrari}
\affiliation{INO-CNR BEC Center and Dipartimento di Fisica, Universit\`a di Trento, 38123 Povo, Italy}
\affiliation{Trento Institute for Fundamental Physics and Applications, INFN,  38123 Povo, Italy}


\date{\today}

\begin{abstract}
We report on the design, construction, and performance of a compact magnetic shield that facilitates a controlled, low-noise environment for experiments with ultracold atomic gases.
The shield was designed to passively attenuate external slowly-varying magnetic fields while allowing for ample optical access.
The geometry, number of layers and choice of materials were optimised using extensive finite-element numerical simulations.
The measured performance of the shield is in good agreement with the simulations.
From measurements of the spin coherence of an ultracold atomic ensemble we demonstrate a remnant field noise of 2.6\,$\mu$G and a suppression of external dc magnetic fields by more than five orders of magnitude.
\end{abstract}


\maketitle 


\section{Introduction}

A low-noise, stable magnetic field is useful in a broad range of scientific fields.
Atom interferometry and microgravity~\cite{kubelka-langeThreelayerMagneticShielding2016,zoestBoseEinsteinCondensationMicrogravity2010},
electron microscopy~\cite{krivanekElectronMicroscopeAberrationcorrected2008}, nuclear magnetic resonance~\cite{mansfieldMultishieldActiveMagnetic1987,taylerInvitedReviewArticle2017}, magnetometry~\cite{bottiNoninvasiveSystemSimultaneous2006,bertoldiNoiseResponseCharacterization2006,bertoldiMagnetoresistiveMagnetometerImproved2005}, and atomic clock~\cite{ludlowOpticalAtomicClocks2015} experiments have all benefited from advances in magnetic field stabilisation.
The dominant magnetic noise in these environments arises from dc magnetic field fluctuations due to geomagnetic fields, other nearby instruments, and magnetised objects.
Specifically, in ultracold atomic
physics~\cite{dedmanActiveCancellationStray2007,zhangCoherentZerofieldMagnetization2015,ottlHybridApparatusBoseEinstein2006} a low-noise, well-controlled magnetic field allows for measuring interaction-driven phenomena that occur at long timescales.
For example, coherently-coupled quantum degenerate mixtures of $^{23}$Na atoms in the miscible $|F,m_F\rangle=|1,\pm 1 \rangle$ states can be used to study sine-Gordon Hamiltonian~\cite{gallemiDecayRelativePhase2019,sonDomainWallsRelative2002,iharaTransverseInstabilityDisintegration2019}.

Passive magnetic shielding is well-suited for isolating an experiment by excluding magnetic fields from a contained volume.
As opposed to active stabilisation~\cite{dedmanActiveCancellationStray2007,merkelMagneticFieldStabilization2019} or dynamical
decoupling~\cite{trypogeorgosSyntheticClockTransitions2018,caiRobustDynamicalDecoupling2012},
it uses materials that have high magnetic permeability $\mu_r$ and so redirect the magnetic flux lines around the enclosed volume.
Different materials have different properties and utilise different shielding mechanisms: high-$\mu_r$ materials screen quasi-dc fields up to a few 100\,Hz by flux-shunting, while highly conductive materials cancel magnetic fields induced by eddy currents oscillating at a few kHz~\cite{sumnerConvectionalMagneticShielding1987,burtOptimalThreelayerCylindrical2002}.

The distortion of the magnetic field depends on the physical parameters of the material, the shield geometry and the frequency of the magnetic source.
For a linear, homogeneous, isotropic, and non-dispersive material the working mechanism is simply understood using Maxwell's equations that relate the flux density $\mathbf{B}$ to the magnetic field $\mathbf{H}$, as $\mathbf{B}(r,t) = \mu_r\mu_0\mathbf{H}(r,t)$,
where $\mu_0$ is the magnetic permeability of vacuum.
In the absence of currents, the requirement that the tangential component of $\mathbf{H}$ and the normal component of $\mathbf{B}$ remain continuous across an interface of materials with different $\mu_r$, the field lines are bent nearly tangentially to the interface~\cite{celozziElectromagneticsShielding2008}.
Although the principle mechanism is simple, an exhaustive design study is still required for optimising all the parameters of the shield.
\begin{figure}[t]
    \centering
    \includegraphics[]{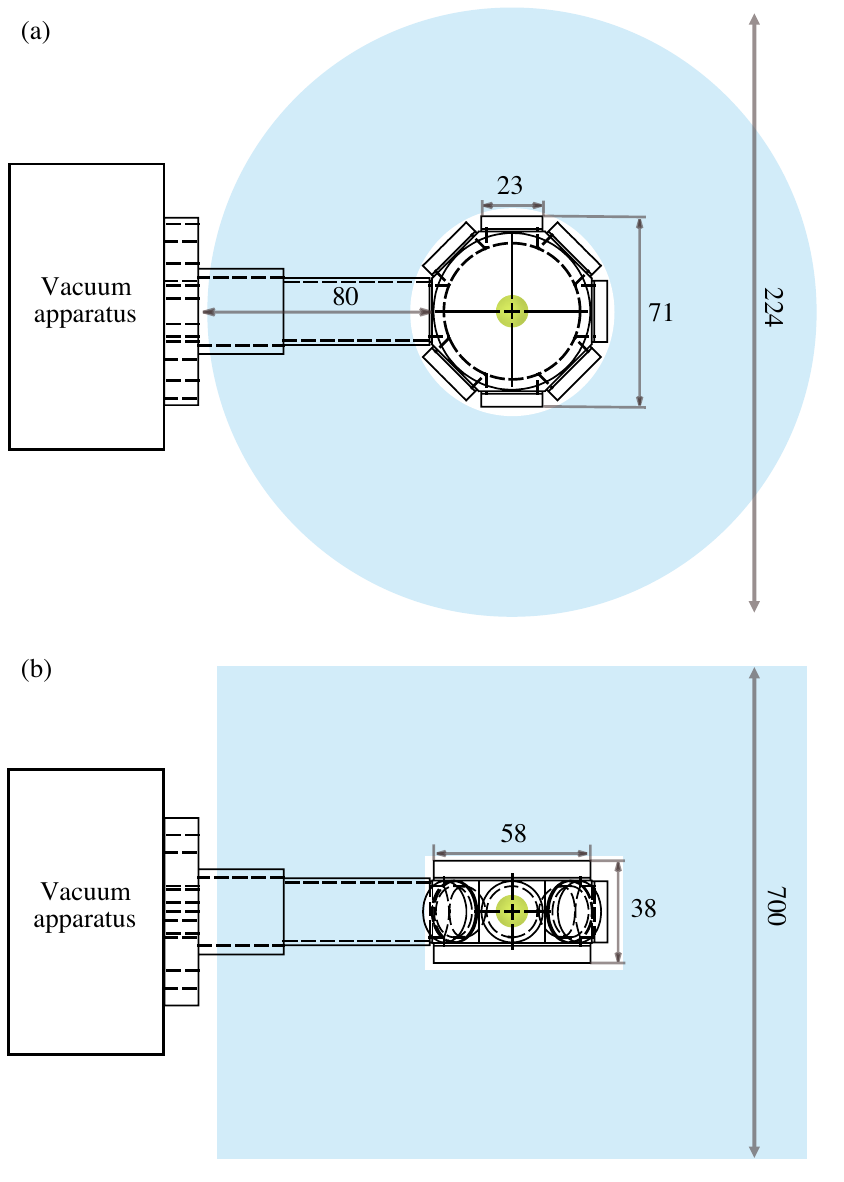}
    \caption{(a) Top and (b) side view of vacuum apparatus.
    The maximum (minimum) dimensions of the shield are 224\,mm (88\,mm) diameter and 700\,mm (50\,mm) length, bound by the vacuum apparatus, the size of the glass cell and the surrounding dc coils, and the presence of additional optics around the glass cell.
    The shield needs to have 10 openings to allow for optical access through all the horizontal/vertical windows of the glass cell and installation around the tube connecting it to the rest of the vacuum apparatus.
    All distances are in millimetres.
    }
    \label{fig:1}
\end{figure}

A compact source of cold $^{23}$Na atoms~\cite{lamporesiCompactHighfluxSource2013,colziSubDopplerCoolingSodium2016}, combined with a hybrid magneto-optical trap~\cite{colziProductionLargeBoseEinstein2018} leave enough space for the construction of a passive magnetic shield able to reduce the fluctuating external magnetic field.
At the same time it is still necessary to produce well-controlled magnetic field inside the shielded volume to fix the quantisation axis and control the energy spacing between atomic states.
The geometric design of the shield makes this possible without compromising its performance.

\section{Design considerations}

The optimisation of the shield design hinges on mainly three aspects: the shape of the shield, the choice of materials, and geometrical constraints of our experiment.
The former sets the trade-off between shielding efficiency and saturation tolerance.
The latter is determined by the geometry of the vacuum apparatus and the need for optical and electrical access.

\subsection{Shape and material}

The ideal shape of a magnetic shield is a sphere or an infinitely long cylinder since sharp corners generally lead to flux leakage.
Precise machining and welding are necessary to minimise this effect.
The overall size of the shield should be as small as possible since the attenuation $A = \mu_r d / 2R$ of an external magnetic field scales as the inverse of the shield radius $R$ at fixed thickness $d$, where $\mu_r$ is the relative permeability of the material.
Multi-layered designs increase the performance of a shield of similar thickness and volume.
The total attenuation is proportional to the product of the attenuation of individual layers.
Each layer sees a reduced magnetic field that is already screened by a previous layer closer to the field source, which makes it unlikely to saturate.

Except for the geometry, the only other parameter that affects the performance of the shield is the magnetic permeability $\mu_r$ which is tied to the choice of the material.
In general ferromagnetic materials have large, non-linear $\mu_r$ that depends on the modulus of $\mathbf H$ and the residual magnetization $\mathbf M$, so that $\mathbf{B}(r,t)/\mu_0 = \mu_r(H)\mathbf H(r,t) + \mathbf M$.
This leads to hysteresis in the response of the material to an external field.
When $\mathbf B$ becomes too high, the magnetic domains of the material are all aligned with the external field, the material saturates and is unable to sustain any higher flux.

Among all ferromagnetic elements and alloys one of the most common materials used for magnetic shielding is $\mu$-metal, which is a magnetically-soft alloy, composed of 80\% nickel, 5\% molybdenum, and iron.
It has $\mu_r=4.7\times 10^5$, which allows it to reach high shielding efficiency but it saturates at relatively small magnetic fields~\footnote{We denote saturation values in units of Tesla, 1\,T = 10$^4$\,G.}, 0.75\,T.
Other materials have a higher saturation value.
For example Supra-50, an alloy composed of 48\% nickel and 52\% iron,  saturates at 1.5\,T, roughly twice the field value of $\mu$-metal but has smaller $\mu_r=2\times 10^5$.

\subsection{Experimental constraints}

The main constraints on the shield geometry are determined by the dimensions of the ultra-high-vacuum cell that contains the atomic gas.
A quartz, octagonal cell is welded on a horizontal tube of $\diameter$25\,mm and 80\,mm length, directly connected to the main vacuum apparatus (see Fig.~\ref{fig:1}).
Each side-face window has $\diameter$23\,mm and a thickness of 4.8\,mm except for the top and bottom windows that are $\diameter$58\,mm and 6.4\,mm thick.
The outer distance between two parallel faces is 71\,mm, while the distance between the center of the cell and the edge of the vacuum apparatus (up to the head of the screws used to fix the flange on which the cell is mounted) is 112\,mm.

The shield comprises multiple pieces so that it can be mounted around the quartz cell.
It cannot be completely hermetic since our experiments require optical access to the atomic sample for laser cooling and trapping~\cite{colziProductionLargeBoseEinstein2018}.
The design requires at least ten apertures of $\diameter$30\,mm so that all directions defined by the cell windows are accessible with standard 25\,mm optics.
Other smaller apertures ($\diameter$10\,mm) are necessary for routing cables that carry current to ac and dc coils inside the shield.
The largest of these is a pair of quadrupole coils that have an inner (outer) diameter of 73\,mm (87\,mm), height of 10\,mm, are 30\,mm apart and can produce a maximum gradient of 200\,G/cm at the location of the atomic ensemble.
The outer diameter of these coils limits the minimum diameter of the shield and the maximum magnetic field they produce needs to be below the saturation threshold of the shield.
The atoms will be exposed to an axial bias magnetic field $\approx 100$\,mG generated by electromagnets, whose modulus is required to be as stable as possible.

These numbers set the minimum and maximum radius of the shield to 44\,mm and 110\,mm respectively, and minimal internal lenght to 50\,mm.
The outer length of the shield is limited by the presence of the optical table supporting the vacuum system, located 350\,mm below the center of the cell.
The overall volume is limited to the dimensions of Fig.~\ref{fig:1}.
With these constraints in mind we proceed with numerical simulations to optimise the design of the shield and assess its performance.

\section{Finite element method simulations}

We used the finite-element method (FEM) as implemented in COMSOL\footnote{COMSOL Multiphysics v. 5.4. www.comsol.com. COMSOL AB, Stockholm, Sweden.} to simulate the behaviour of the shield for different geometrical configurations.
In a series of simulations we optimised the number of layers of the shield, their thickness, and inter-layer distance considering both the suppression of external magnetic fields and the saturation of the inner layer due to fields produced by our dc coils.

\subsection{Number of layers}

\begin{figure}[t]
    \centering
    \includegraphics[width=\columnwidth]{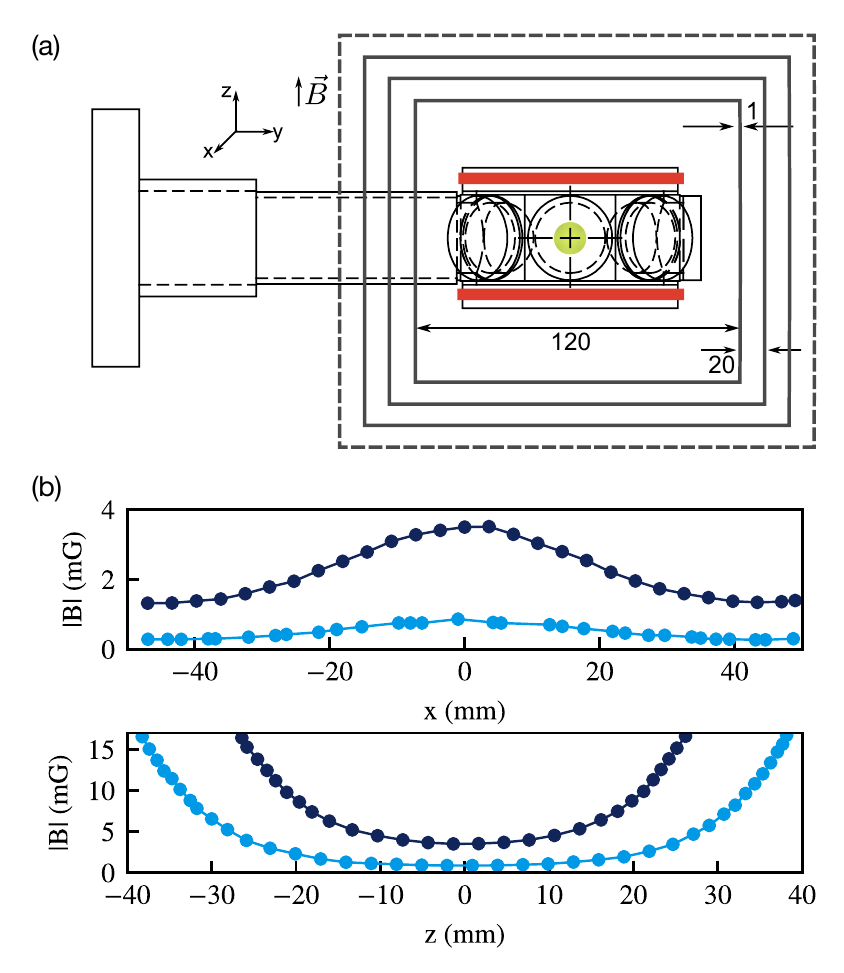}
    \caption{(a) Shield geometry used to investigate the difference between 3 and 4 layers. The inner layer was a cylinder with diameter and length equal to 120\,mm and each subsequent layer was increased by 40\,mm both in diameter and length. All layers were 1\,mm thick.
    The applied external field was 1\,G directed along $z$.
    (b) Magnetic flux density along $x$ (top) and $z$ (bottom) in the central region of the shield. Using more layers leads to a higher suppression of the external field along both directions.}
    \label{fig:2}
\end{figure}
We compared the attenuation of an external field directed along the vertical and horizontal axes, of a 3-layer and a 4-layer shield.
For this simulation we used a simplified geometry composed of three or four concentrical cylinders, which are exposed to an external uniform magnetic field of 1\,G.
The innermost cylinder had an internal radius of 60\,mm, length of 120\,mm and thickness of 1\,mm.
Each subsequent layer was increased by 40\,mm both in diameter and length.
In order to save computational time and memory, the magnetic permeability of the material used was set to $\mu_r=4\times10^4$, a conservative value with respect to the actual response of $\mu$-metal.
Both geometries attenuate the external field by three orders of magnitude with the 4-layer shield outperforming the 3-layer shield by more than a factor of 3 (Fig.~\ref{fig:2}).
It also leads to a more uniform field, especially along the horizontal plane.

\subsection{Saturation}

\begin{figure}[t]
    \centering
    \includegraphics[width=\columnwidth]{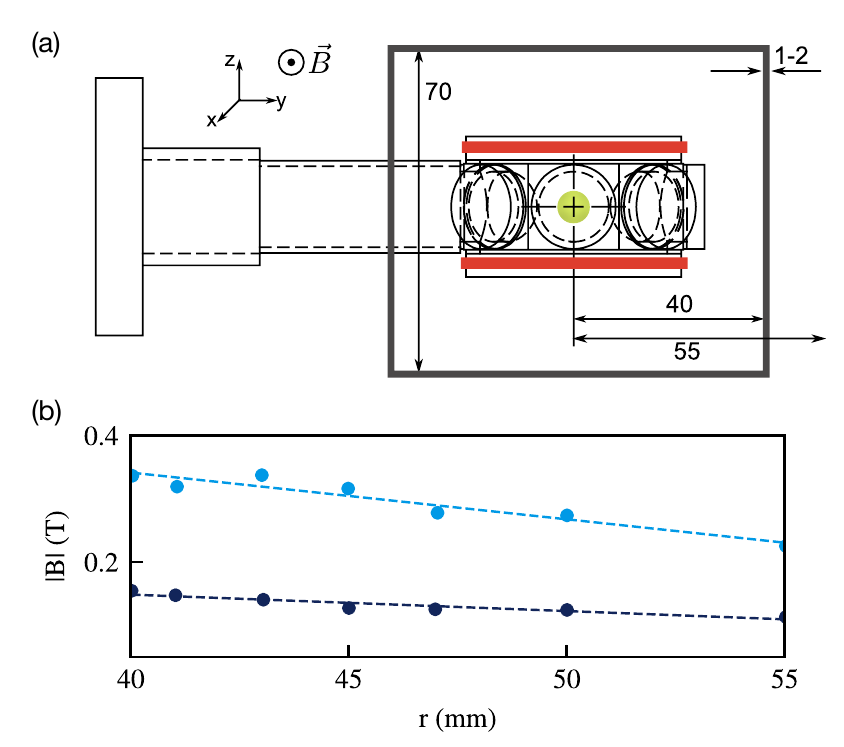}
    \caption{(a) Geometry used to investigate the effect of the non-uniform field produced by the dc coils on the shield. The shield has a length of 70\,mm, a varying radius from 40\,mm to 55\,mm and a thickness of 1\,mm (light) or 2\,mm (dark).
    (b) Maximum value of the magnetic flux inside the shield layer as a function of its radius. The field decreases by a factor of two for the thicker layer. The dashed lines are linear fits to the data showing a decrease of 75\,G/mm for the 1\,mm thick layer and 26\,G/mm for the 2\,mm thick layer.}
    \label{fig:3}
\end{figure}

We then simulated the effect of the internal magnetic fields produced by the dc coils on the saturation of the shield material.
We considered the magnetic field produced by a pair of coils placed above and below the two upper and lower windows of the glass cell with a mean radius and a relative distance of 31\,mm, producing a gradient field of 100\,G/cm.
We modelled a single-cylinder $\mu$-metal shield with a length of 70\,mm, radius of 40\,mm and variable thickness of 1\,mm or 2\,mm surrounding the coils.
Increasing the thickness or the distance between the shield and the magnetic field source lowers the magnetic flux in the bulk of the material.
For a radius of 40\,mm, the maximum field value on the xy-plane is 0.33(1)\,T versus 0.15(1)\,T for a thickness of 1\,mm and 2\,mm respectively.
The flux is more than halved when the thickness is doubled and the decay rate of the magnetic flux as a function of the radius is larger for smaller thickness (Fig.~\ref{fig:3}).

The maximum field the shield experiences lies away from the xy-plane, near the edge of the dc coils and is 0.5\,T, close to the saturation threshold of $\mu$-metal.
Due to this, we proceeded with a more realistic simulation of a single-cylinder shield made of Supra-50, that includes all the apertures needed.
The shield was 2.5\,mm thick, had a length of 211\,mm and a radius of 54\,mm.
For a gradient of 50\,G/cm the maximum field generated on the shield is 250\,G which is well below the saturation threshold of Supra-50.

\subsection{Axial attenuation}

The optimal spacing of the layers of the shield is attained when the radius of each layer is double that of the previous one~\cite{burtOptimalThreelayerCylindrical2002}.
This leads to prohibitively large shield designs that cannot be accommodated in our experiment due to space constraints.
We still optimised the axial inter-layer distance using a 4-layer shield with an inner-layer radius of 57\,mm, length of 84\,mm, radial inter-layer distance of 20\,mm, and an external magnetic field of 1\,G along the $x$- or $z$-axis.
We investigated the role of the length of the innermost layer including all the openings.
Increasing the length of the innermost layer in steps of 10\,mm while keeping the axial inter-layer distance fixed at 10\,mm increases the shielding efficiency, and leads to a more uniform residual magnetic field in the region of interest.
The absolute value of the residual magnetic field at the centre of the shield decreases from 416\,$\mu$G to 154\,$\mu$G when the length of the shield increases from 84\,mm to 124\,mm.

Similarly, keeping the length of the innermost layer fixed at 84\,mm and varying the axial inter-layer distance from 10\,mm to 30\,mm the residual field goes from 416\,$\mu$G to 43\,$\mu$G.
The results of these simulations are shown in Table~\ref{tab:1}.
By fixing the inter-layer distance to 10\,mm and increasing the length of the innermost layer, we were able to obtain good shielding performance along the axial direction.
\begin{table}[ht]
  \caption{\label{tab:1} Residual magnetic field values for a 4-layer shield of varying length and inter-layer distance. The inter-layer distance (length) was fixed at 10\,mm (84\,mm) when varying the length (inter-layer distance). Longer designs with larger spacing between the layers lead to better performance.}
  \begin{ruledtabular}
    \begin{tabular}{cccccc}
      length (mm)& 84 & 104 & 124 & 84 & 84 \\
      inter-layer distance (mm) & 10 & 10 & 10 & 20 & 30 \\
      $\vert B\vert$ ($\mu$G) & 416 & 197 & 154 & 69 & 43 \\
    \end{tabular}
  \end{ruledtabular}
\end{table}

\section{Implementation and characterisation}

\subsection{Implementation}

\begin{figure}[t]
    \centering
    \includegraphics[]{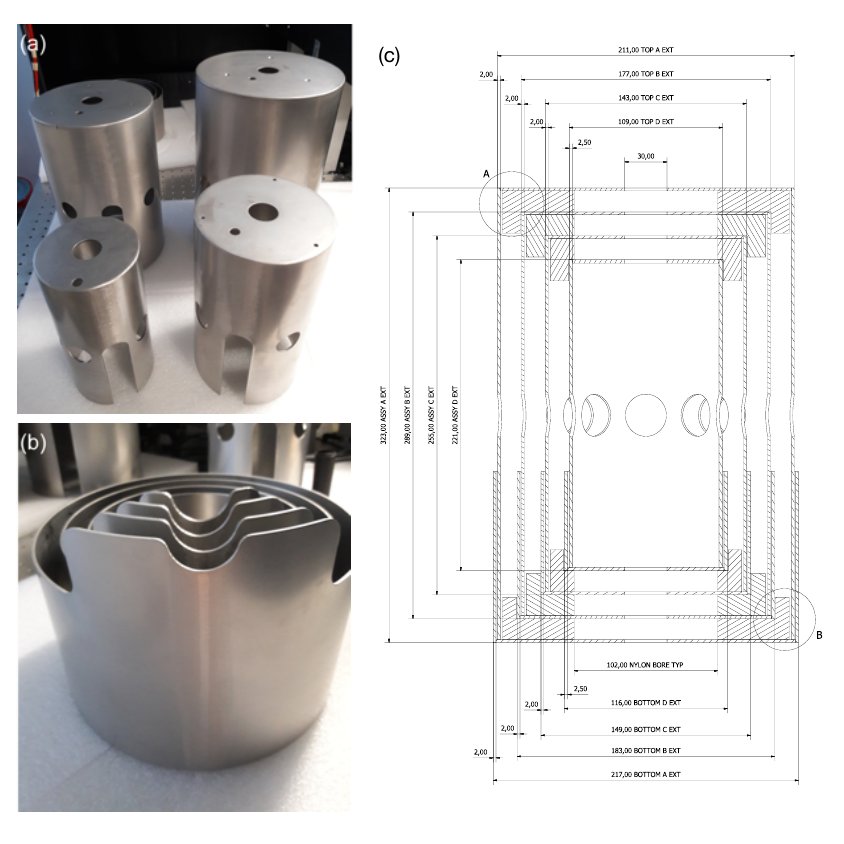}
    \caption{(a) Picture of the four layers that comprise the upper half of the shield. The lower left layer is made out of Supra-50 and the rest from $\mu$-metal.
    (b) The four bottom layers stacked into their final configuration. The layers are spaced using a set of nylon supports.
    (c) Technical drawing of the final shield arrangement showing all the relevant dimensions and openings. The shaded areas at the top and bottom are the nylon supports.}
    \label{fig:4}
\end{figure}

Based on the results of our simulations we finalised the design of the shield to that shown in Fig.~\ref{fig:4}.
The shield is composed of four layers of different material and sizes.
Its volume is 12 liters which makes it much more compact than existing designs~\cite{xuUltralowNoiseMagnetic2019}.
The innermost layer is a 2.5\,mm-thick Supra-50 layer, while the outer layers are made out of 2\,mm-thick $\mu$-metal.
Each layer is composed of two top/bottom pieces that are stacked on each other.
The internal radius of the bottom piece is equal to the external radius of the top piece plus a 1\,mm clearance.
This structure is modular and easy to assemble and disassemble.
We used two sets of nylon supports to fix the spacing between subsequent layers (both top and bottom) along the axial and radial direction during assembly.
The shield has ten $\diameter$30\,mm and two $\diameter$10\,mm openings, which are used for optical and cable access respectively.
An additional set of three holes of $\diameter$4\,mm are present at the top and at the bottom of the outer layers to connect to the mounting pieces.
Conclusive simulations accounting for the non-linear response and the saturation limit of the layers predict for this geometry an overall attenuation of an external magnetic field by a factor of $3\times 10^5$ (Fig.~\ref{fig:5}), and that the magnetic field produced by the dc coils leaves the inner-shield layer effectively unmagnetised.

\subsection{Characterisation}

We tested the performance of the shield before mounting it on the experimental apparatus, using a solenoid with a rectangular cross-section 57\,cm$\times39$\,cm and length of 1.5\,m, large enough to encompass the whole shield and provide a uniform magnetic field throughout its volume.
The solenoid was made out of 138 windings of 0.8\,mm copper wire, spaced by a mean distance of 10\,mm and produced a field of 1.28\,G/A.
We profiled the axial field of the solenoid using differential $\pm$0.5\,A measurements and found it to be uniform to within 6\%.
We measured the attenuation of the solenoid magnetic field due to the presence of the shield layers using a Mag-13MCL100 low-noise 100\,$\mu$T field sensor from Bartington Instruments.
\begin{figure}[ht!]
    \centering
    \includegraphics[]{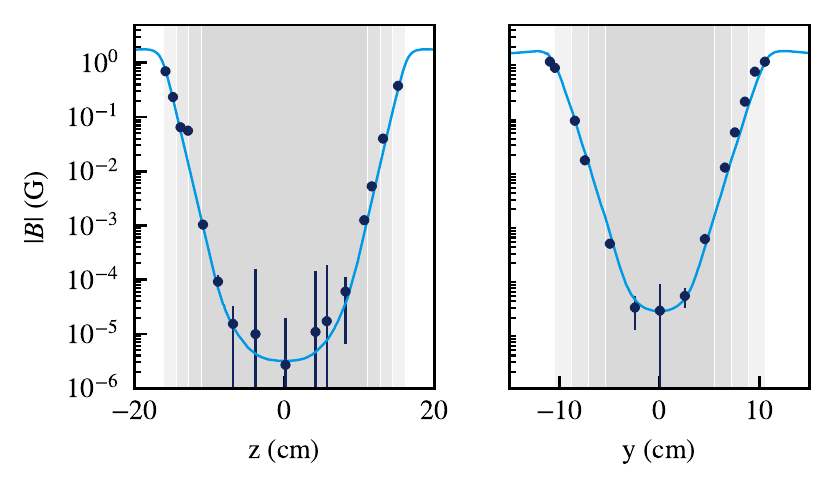}
    \caption{Measurement of the magnetic field suppression along the axial and radial directions (points) compared to the simulated data (lines).
    The error on the horizontal axis is $\pm$3\,mm which is less than the width of the points.
    On the vertical axes we assume a systematic error of 6\% in addition to the resolution of the magnetometer.
    The shaded vertical regions corresponds to the shield layers.}
    \label{fig:5}
\end{figure}
The measurements are shown in Fig.~\ref{fig:5} along with the simulated field inside the shield.
The remnant field inside the shield is less than 30\,$\mu$G and 3\,$\mu$G along the radial and axial directions respectively for a 1\,G external field which represents a suppression of roughly 6 orders of magnitude.

After the shield was positioned in place around the vacuum cell, we tested its performance using ultracold atoms as magnetic field sensors.
We measured the fluctuations of the magnetic field using a, microwave two-photon transition between $\ket{1, -1}$ and $\ket{1, 1}$ of $^{23}$Na ultracold atoms in a magnetic field with a Larmor frequency of 180\,kHz.
After finding the resonant frequency between the two states, we use side-of-fringe pulses to prepare an equal superposition of the two states.
We measure the atomic magnetisation fluctuations over time, which is sensitive to the Larmor-frequency shift due to the linear Zeeman effect.
Using an independent calibration of the Rabi frequency we compute the magnetic field fluctuations from the magnetisation.

\begin{figure}[t!]
    \centering
    \includegraphics[]{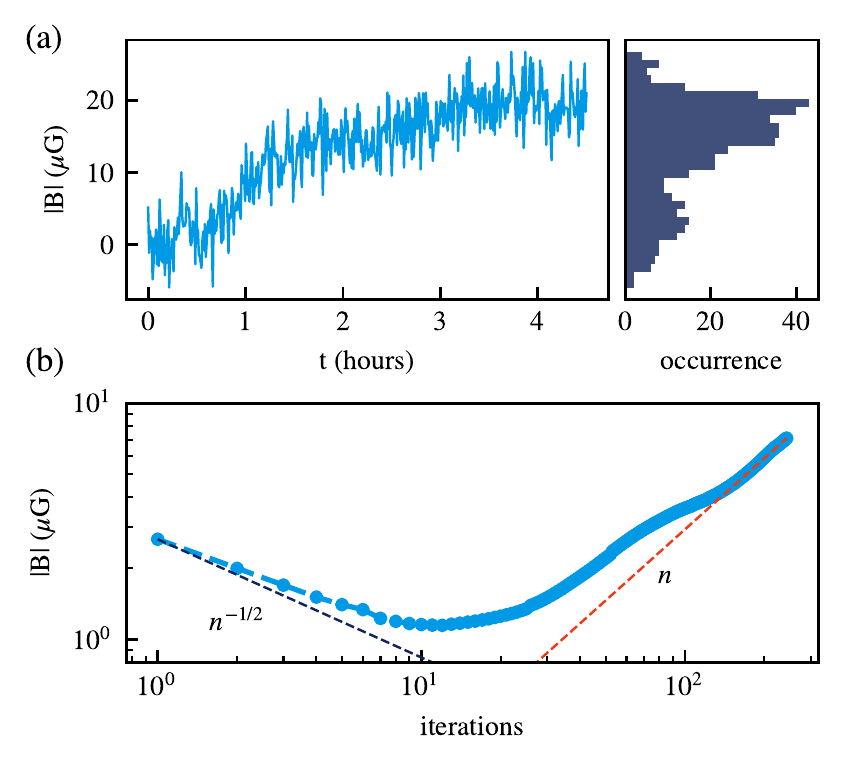}
    \caption{(a) Calculated magnetic field from atomic spectroscopy measurements over a period of 4.5 hours (left).
    The field drifts by about 20\,$\mu$G in that time while the shot-to-shot fluctuations are 2.6\,$\mu$G.
    Histogram of the distribution of the biased magnetic field fluctuations (right).
    (b) The Allan deviation of the field over the number of experimental iterations $n$ shows that the magnetic field stability is as low as 1.1\,$\mu$G after 500\,s.
    At short times the noise is dominated by shot-to-shot fluctuations $\propto n^{-1/2}$ while at long times the drift term $\propto n$ is larger.
    The cycle time of the experiment was 32\,sec and the interrogation time of the magnetic field was 2.6\,ms.}
    \label{fig:6}
\end{figure}

Figure~\ref{fig:6} shows the fluctuations of the magnetic field resonance over a period of 4.5 hours.
The biased fluctuations are clearly dominated by a slow drift of the order of 20\,$\mu$G, which we also verified by spectroscopical measurements before and after this stability measurement.

The Allan deviation (Fig.~\ref{fig:6}b) gives further insight into the nature of the fluctuations.
Each point in the graph corresponds to a single realisation of the experiment with a cycle time of 32\,sec probing the field value for 2.6\,ms.
From the first few points we infer a standard deviation of the unbiased fluctuations of 2.6\,$\mu$G.
At short times the noise is limited by shot-to-shot and technical noise.
After an integration time of 500\,sec the minimum deviation of the field is as low as 1.1\,$\mu$G.
At larger times the Allan deviation is dominated by the drift term but still remains below 7\,$\mu$G.
The drift can be corrected with low-bandwidth feedforward loop to eliminate this term~\cite{dedmanActiveCancellationStray2007,xuUltralowNoiseMagnetic2019}.
The ambient magnetic field around the shield had an r.m.s. noise of $220\,\mu$G as measured by the magnetometer.
This was mainly due to the 50\,Hz line-noise and its first odd harmonic.
Without the shield in place we also observed abrupt magnetic field shifts of a few mG induced by the presence/absence of nearby magnetised objects up to a few meters away\footnote{The magnetised object was actually a car that was parked outside the window of our lab, so we affectionately refer to this measurement as `car-acterisation' of the magnetic field}.
We measured a contribution of the line-noise, by scanning the time at which the field measurement was taken, to be roughly equal to the shot-to-shot fluctuations.
This term was not present in the measurements of Fig.~\ref{fig:6} since the experiment was phase-locked to the line frequency.

At the beginning of each sequence we used a degaussing ramp~\cite{thielDemagnetizationMagneticallyShielded2007} to account for the magnetic hysteresis of the shield.
The degaussing ramp was comprised of 10 current pulses of alternating polarity whose amplitude diminished exponentially from 20\,A to 0.2\,A over 1.5 seconds and produced a field of 1.3\,G/A at the location of the atoms.
We observed that after changing the magnetic field environment inside the shield, it takes about 20 repetitions of the experiment for the magnetic field to relax to its new value due to residual hysteresis of the shield.

\section{Conclusions}

Passive stabilisation of the ambient magnetic field using a precisely engineered ferromagnetic shield works extremely well for atomic physics experiments.
Based on our extensive design study, we presented the implementation and characterisation of a compact magnetic shield that reduces the low-frequency noise in our laboratory by six orders of magnitude to the level of a few $\mu$G.
This will allow for a new generation of experiments to be performed.
For example, experiments where the coherence of the internal spin dynamics of a Rabi-coupled atomic system is longer than the typical timescales of many-body dynamics~\cite{gallemiDecayRelativePhase2019,tylutkiConfinementPrecessionVortex2016} and of the order of a few Hz, or that require the manipulation of atomic interactions through rf-induced Feshbach resonances~\cite{papoularMicrowaveinducedFanoFeshbachResonances2010} will be possible.

%
%

%

\begin{acknowledgments}
We acknowledge discussions and suggestions from MuShield Inc. and Amuneal Manufacturing. We thank Magnetic Shield Ltd. for collaborating on the design and manufacturing of the shield. We acknowledge fruitful discussions with Immanuel Bloch and Christophe Salomon.

This work was supported by the Provincia Autonoma di Trento and the Istituto Nazionale di Fisica Nucleare under the FIS$\hbar$ project.
We acknowledge received funding from the project NAQUAS of QuantERA ERA-NET Cofund in Quantum Technologies (Grant Agreement N. 731473) implemented within the European Union's Horizon 2020 Programme.
\end{acknowledgments}

\bibliography{magnetic_shield}

\end{document}